\documentclass[reviewcopy]{elsart}
\usepackage{graphicx}
\usepackage{amssymb}

\begin{document}

\begin{frontmatter}



\title{Influence of electromagnetic interferences on the gravimetric sensitivity of surface acoustic waveguides.}


\author[LLN]{L. Francis}\ead{francis@pcpm.ucl.ac.be},
\author[HEV]{J.-M. Friedt},
\author[HEV]{R. De Palma},
\author[LLN]{P. Bertrand} and
\author[HEV]{A. Campitelli}

\address[LLN]{PCPM, Universit{\'e} catholique de Louvain, Croix du Sud 1, B-1348 Louvain-la-Neuve, Belgium}
\address[HEV]{Biosensors Group, IMEC, Kapeldreef 75, B-3001 Leuven, Belgium}

\begin{abstract}
Surface acoustic waveguides are increasing in interest for
(bio)chemical detection. The surface mass modification leads to
measurable changes in the propagation properties of the waveguide.
Among a wide variety of waveguides, Love mode has been
investigated because of its high gravimetric sensitivity. The
acoustic signal launched and detected in the waveguide by
electrical transducers is accompanied by an electromagnetic wave;
the interaction of the two signals, easily enhanced by the open
structure of the sensor, creates interference patterns in the
transfer function of the sensor. The influence of these
interferences on the gravimetric sensitivity is presented, whereby
the structure of the entire sensor is modelled. We show that
electromagnetic interferences generate an error in the
experimental value of the sensitivity. This error is different for
the open and the closed loop configurations of the sensor. The
theoretical approach is completed by the experimentation of an
actual Love mode sensor operated under liquid in open loop
configuration. The experiment indicates that the interaction
depends on the frequency and the mass modifications.
\end{abstract}


\end{frontmatter}


\section{Introduction}
\label{introduction}

Acoustic waves guided by the surface of solid structures form
waveguides used as delay lines and filters in telecommunications
\cite{Campbell89}. Waveguides support different modes with
specific strain and stress fields \cite{Auld73b}. The acoustic
velocity of each mode depends on different intrinsic and extrinsic
parameters such as the mechanical properties of the materials, the
temperature or the applied pressure. Waveguides are used as
sensors when the velocity change is linked to environmental
changes. For gravimetric sensors, the outer surface of the
waveguide is exposed to mass changes. Due to the confinement of
the acoustic wave energy close to the surface, these sensors are
well suited for (bio)chemical sensors operating in gas or liquid
environments. Among a wide variety of waveguides used for that
purpose, Love mode sensors have attracted an increasing interest
during the last decade \cite{KKZ03,Tamarin03,Gizeli97,Harding97}.
A Love mode is guided by a solid overlayer deposited on top of a
substrate material. The usual substrates are piezoelectric
materials like quartz, lithium tantalate and lithium niobate
\cite{Herrmann01}. Associated to specific crystal cut of these
substrates, the Love mode presents a shear-horizontal polarization
that makes it ideal for sensing in liquid media.

Current research in Love mode sensors concerns the guiding
materials in order to obtain a high gravimetric sensitivity.
Typical materials under investigations are dielectrics like
silicon dioxide and polymers, and more recently semiconductors
with piezoelectric properties like zinc oxide
\cite{Chu03,Rasmusson01,Harding01,Du98}. Although the dispersion
relation for Love mode is well set and the dependence of the
gravimetric sensitivity of the liquid loaded sensor to the
overlayer thickness has been thoroughly investigated
\cite{McHale02,Jakoby98,Jakoby97,Wang94}, little has been devoted
to study the role played by the structure of the sensor and the
transducers.

In this paper, we investigate the role played by the structure of
the sensor and the interferences between the acoustic and the
electromagnetic waves on the gravimetric sensitivity. In the first
part, we present a model of the transfer function including the
influence of electromagnetic interferences. In the second part, we
show how these interferences modify the gravimetric sensitivity in
open and closed loop configurations of the sensor. Finally, these
effects are illustrated experimentally on a Love mode sensor.

\section{Modelling}
\label{modelling}

Waveguide sensors consist of a transducing part and a sensing
part. The transducing part includes the generation and the
reception of acoustic signals and their interfacing to an
electrical instrumentation. The most common transducers are the
widespread interdigital transducers (IDTs) on piezoelectric
substrates introduced by White and Voltmer in 1965 \cite{White65}.
Although the transducing part can be involved in the sensing part,
practical sensing is confined to the spacing between the
transducers. This confinement takes especially place when liquids
are involved since these produce large and unwanted capacitive
coupling between input and output electrical transducers. This
coupling dramatically deteriorates the transfer function and is an
important issue for the instrumentation of the sensors.

The sensor is a delay line formed by the transducers and the
distance separating them. Each transducer is identified to its
midpoint. The distance between the midpoints is $L$. The sensing
part is located between the transducers and covers a total length
$D$ so that $D \le L$. The guided mode propagates with a phase
velocity $V=\omega / k$, where $\omega=2\pi f$ is the angular
frequency and $k=2\pi /\lambda$ is the wavenumber. The waveguide
is dispersive when the group velocity $V_g=\d \omega/ \d k$
differs from the phase velocity.

The velocity is a function of the frequency and of the surface
density $\sigma=M/A$ for a rigidly bound mass $M$ per surface area
$A$. For an uniformly distributed mass, the surface density is
rewritten in terms of material density $\rho$ and thickness $d$ by
$\sigma=\rho d$. The phase velocity for an initial and constant
mass $\sigma_0$ is denoted $V_0$, and the group velocity $V_{g0}$.
In the sensing part, the phase velocity is $V$ and the group
velocity $V_g$. According to this model, the transit time $\tau$
of this delay line is given by:
\begin{equation}\label{eq:tau1}
    \tau=\frac{D}{V}+\frac{L-D}{V_0}.
\end{equation}

Electromagnetic interferences are due to the cross-talk between
the IDTs. The electromagnetic wave (EM) emitted by the input
transducer travels much faster than the acoustic wave and
therefore is detected at the output transducer without noticeable
delay. At the output transducer, the two kinds of waves interact
with an amplitude ratio, denoted by $\alpha$, that creates
interference patterns in the transfer function $H(\omega)$ of the
delay line. The transfer function itself is given by the ratio of
the output to the input voltages. The transfer function with
electromagnetic interferences is modelled by the following
equation:
\begin{equation} \label{eq:transfertA}
H(\omega)=\underbrace{H_0(\omega)\exp(-\mathrm{i}\omega
\tau)}_{\mathrm{delay\ line}} +\underbrace{\alpha
H_0(\omega)}_{\mathrm{EM\ coupling}}.
\end{equation}
The transfer function $H_0(\omega)$ is associated to the design of
the transducers. The total transfer function can be rewritten as
$H(\omega)=\left\|H(\omega)\right\| \exp\left(\rm{i}\phi\right)$
where expressions for the amplitude $\left\|H(\omega)\right\|$ and
the phase $\phi$ are obtained with help of complex algebra:
\begin{eqnarray} \label{eq:H}
  \left\|H(\omega)\right\|&=& \left\|H_0(\omega)\right\|\left\|\sqrt{1+2\alpha
  \cos(\omega\tau)+\alpha^2}\right\|; \\ \label{eq:phi}
  \phi&=&\phi_0-\arctan{\left(\frac{\sin(\omega\tau)}{\alpha +\cos(\omega\tau)}\right)}.
\end{eqnarray}
The phase $\phi_0$ corresponds to the packaging of the sensor and
is due to different aspects linked to the instrumentation. It will
be assumed independent of the frequency and of the sensing event.
The relations (\ref{eq:H}) and (\ref{eq:phi}) are the sources of
ripples in the transfer function at the ripple frequency $\Delta
\omega = 2\pi/\tau$. The relative amplitude peak to peak of the
perturbation on the amplitude has a maximum effect (in dB) equals
to $40\log[(1+\alpha)/(1-\alpha)]$. The amplitude (in dB and
normalized to have $\|H_0(\omega)\|=1$) and the phase (in radians)
as a function of the frequency are simulated in
Figures~\ref{fig:phasealpha0} to \ref{fig:phasealpha3} for
different values of $\alpha$.

Under the influence of the interferences, the phase has different
behaviors function of $\alpha$:
\begin{enumerate}
    \item when $\alpha=0$ (no interferences), the phase is linear with
the frequency and has a periodicity equal to $2\pi$
(Fig.\ref{fig:phasealpha0});
    \item when $\alpha <1$, the phase is deformed but has still a
periodicity equal to $2\pi$ (Fig.~\ref{fig:phasealpha1});
    \item when $\alpha =1$, the phase has a periodicity equal to $\pi$
(Fig.~\ref{fig:phasealpha2});
    \item when $\alpha >1$, the periodicity is lower than $\pi$
    (Fig.~\ref{fig:phasealpha3});
    \item when $\alpha \rightarrow \infty$, the phase is not periodic anymore and
its value tends to $\phi_0$.
\end{enumerate}
This specific behavior of the phase has to be considered for the
evaluation of the gravimetric sensitivity.

\section{Gravimetric sensitivity}
\label{gravimetricsensitivity}

Changes in the boundary condition of the waveguide due to the
sensing event modify phase and group velocities. As consequence,
the transit time of the delay line and the phase of the transfer
function are modified. The sensing event is quantified by
recording the phase shift at a fixed frequency (open loop
configuration) or the frequency shift at a fixed phase (closed
loop configuration). This quantification gives rise to the concept
of sensitivity. The sensitivity is the most important parameter in
design, calibration and applications of acoustic waveguide
sensors. Its measurement must be carefully addressed in order to
extract the intrinsic properties of the sensor.

\subsection{Sensitivity definitions}
The {\em gravimetric sensitivity} $S_V$ is defined by the change
of phase velocity as a function of the surface density change at a
constant frequency. Its mathematical expression is given by
Ref.~\cite{Jakoby97}:
\begin{equation}\label{eq:Sv}
    S_V= \left. \frac{1}{V} \frac{\partial V}{\partial
    \sigma}\right|_\omega.
\end{equation}

The definition reflects the velocity change in the sensing area
only while outside this area the velocity remains unmodified. The
expression is general because the initial velocity $V$ of the
sensing part does not need to be equal to $V_0$; this situation
occurs in practical situations where the sensing part has a
selective coating with its own mechanical properties, leading to a
difference between $V$ and $V_0$.

To link the gravimetric sensitivity (caused by the unknown
velocity shift) to the experimental values of phase and frequency
shifts, we introduce two additional definitions related to the
open and the close loop configurations, respectively. The {\em
phase sensitivity} $S_\phi$ is defined by
\begin{equation}\label{eq:SphiA}
 S_\phi = \frac{1}{kD}\frac{\d \phi}{\d \sigma},
\end{equation}
and the {\em frequency sensitivity} $S_\omega$ is defined by
\begin{equation}\label{eq:Somega}
    S_\omega = \frac{1}{\omega} \frac{\d \omega}{\d \sigma}.
\end{equation}

\subsection{Phase differentials without interferences}

In order to point clearly the effects of the electromagnetic
interferences on the different sensitivities presented in the
previous section, we calculate the phase differentials in the
ideal case of no interferences. For that case, the phase of the
transfer function is a function of the frequency and the velocity,
itself function of the frequency and the surface density:
\begin{eqnarray}\label{ap:eq:A}
\phi(\omega,V(\omega,\sigma)) &=& -\omega \tau\\
&=&-\omega \left( \frac{D}{V} + \frac{L-D}{V_0} \right).
\end{eqnarray}
Therefore, its total differential is:
\begin{eqnarray} \label{ap:eq:Ba}
\d \phi & = &  \left. \frac{\partial \phi}{\partial
    \omega}\right|_\sigma \d \omega +  \left. \frac{\partial \phi}
    {\partial \sigma}\right|_\omega \d \sigma; \\ \label{ap:eq:B}
    & = & \left( \left. \frac{\partial \phi}{\partial \omega} \right|_V +
    \left. \frac{\partial \phi}{\partial V} \right|_\omega \left.
\frac{\partial V}{\partial \omega} \right|_\sigma \right) \d
\omega + \left. \frac{\partial \phi}{\partial V} \right|_\omega
\left. \frac{\partial V}{\partial \sigma} \right|_\omega \d
\sigma.
\end{eqnarray}

The derivative of the phase velocity as a function of the
frequency comes from the definitions of phase and group
velocities; at constant surface density we have from
Ref.~\cite{McHale02}:
\begin{equation} \label{ap:eq:C}
\left. \frac{\partial V}{\partial \omega}\right|_\sigma = k^{-1}
\left( 1 - \frac{V}{V_g} \right).
\end{equation}

The other partial differentials are obtained by differentiation of
Eq.~(\ref{ap:eq:A}):
\begin{eqnarray} \label{ap:eq:dphidomV}
\left. \frac{\partial \phi}{\partial
\omega}\right|_V & = & -\tau \\
\label{ap:eq:D} \left. \frac{\partial \phi}{\partial
V}\right|_\omega &=& \frac{\omega D}{V^2}; \\
\label{ap:eq:dphidomsigma} \left. \frac{\partial \phi}{\partial
\omega}\right|_\sigma &=& -\tau - \omega \left. \frac{\partial
\tau}{\partial \omega} \right|_\sigma.
\end{eqnarray}
The partial differential of $\tau$ is given by
\begin{equation}\label{eq:tauprime}
    \left. \frac{\partial \tau}{\partial \omega} \right|_\sigma =
    -\left(\frac{D(V_g-V)}{\omega VV_g} + \frac{(L-D)(V_{g0}-V_0)}
    {\omega V_0 V_{g0}} \right)
\end{equation}
and introduced in Eq.~(\ref{ap:eq:dphidomsigma}), it simplifies
the expression in the form:
\begin{equation} \label{ap:eq:E}
\left. \frac{\partial \phi}{\partial \omega}\right|_\sigma =
-\left( \frac{D}{V_g} + \frac{L-D}{V_{g0}} \right).
\end{equation}

In the case without dispersion, we have an equality between the
partial derivative of the phase with respect to the frequency at
constant surface density or at constant velocity as given by
Eq.~(\ref{eq:tau1}):
\begin{equation}\label{ap:eq:F}
\left. \frac{\partial \phi}{\partial \omega} \right|_\sigma =
\left. \frac{\partial \phi}{\partial \omega} \right|_V   = -\tau.
\end{equation}

\subsection{Open loop configuration}
In the open loop configuration, the input transducer is excited at
a given frequency while the phase difference between output and
input transducers is recorded. This configuration with a constant
frequency has $\d \omega=0$ in Eq.~(\ref{ap:eq:B}); related phase
variations caused by surface density variations are obtained by
\begin{eqnarray}\label{eq:Svopenloop}
    \frac{\d \phi}{\d \sigma} &=&\left. \frac{\partial \phi}{\partial V}
    \right|_\omega \left. \frac{\partial V}{\partial \sigma}
    \right|_\omega\\ \label{eq:Svopenloop2}
    &=&\left. \frac{\partial \phi}{\partial V} \right|_\omega
    VS_V.
\end{eqnarray}

In the absence of interferences, phase variations obtained
experimentally are directly linked to velocity changes by the
product $kD$ involving the geometry of the sensor as seen by
replacing Eq.~(\ref{ap:eq:D}) in Eq.~(\ref{eq:Svopenloop2}):
\begin{equation}\label{eq:ap:dphidsigma}
\frac{\d \phi}{\d \sigma} = kDS_V.
\end{equation}
In other words: $S_\phi=S_V$ when there are no interferences. In a
first approximation $k$ is assumed equal to $k_0$ determined by
the periodicity of the interdigitated electrodes in the
transducer. This assumption is valid as long as the phase shift is
evaluated close to the central frequency $\omega_0=V_0k_0$ and for
waveguides with low dispersion. The wavelength is only known when
the sensing part extends to the transducers ($D=L$). In that case,
the transfer function of the transducers is modified as well by
the velocity changes. In practice the value of the sensitivity is
slightly underestimated to its exact value since $k \le k_0$, the
error being of the order of $5\%$.

In the case where interferences occur, the partial differential of
$\phi$ with respect to the velocity is obtained by differentiation
of Eq.~(\ref{eq:phi}):
\begin{equation} \label{eq:dphidVatom}
\left. \frac{\partial \phi}{\partial V} \right|_\omega=
\left(\frac{1+\alpha \cos(\omega \tau)}{1+2\alpha\cos(\omega
\tau)+\alpha^2} \right) \frac{\omega D}{V^2},
\end{equation}
and the phase sensitivity is obtained by combining the latter
equation with Eq.~(\ref{eq:Svopenloop2}):
\begin{equation}\label{eq:phaseSinterf}
S_\phi = \left( \frac{1+\alpha \cos(\omega \tau)}{1+2\alpha
\cos(\omega \tau) + \alpha^2}\right) S_V.
\end{equation}
The influence of electromagnetic interferences on the phase
sensitivity is simulated in Figure~\ref{fig:constantfreq} versus
the relative frequency for different values of $\alpha$. The phase
sensitivity is always different compared to the gravimetric
sensitivity. For the threshold value $\alpha=1$, the phase
sensitivity is half of the gravimetric sensitivity; for higher
values of $\alpha$, the phase sensitivity is always underestimated
to the gravimetric sensitivity.

\subsection{Closed loop configuration}
In the closed loop configuration, the frequency is recorded while
a feedback loop keeps the phase difference between output and
input transducers constant.
The configuration at constant phase has $\d \phi =0$, the
variation of the frequency as a function of the mass change is
given by introducing this condition in Eq.~(\ref{ap:eq:Ba}):
\begin{equation}\label{eq:phaseOmSigma2}
    \frac{\d \omega}{\d \sigma}= -\left. \left({\left. \frac{\partial \phi}
    {\partial \sigma}\right|_\omega}\right) \right/ \left( {\left. \frac{\partial \phi}
    {\partial \omega}\right|_\sigma} \right).
\end{equation}
The upper term is replaced by Eq.~(\ref{eq:Svopenloop2}). The
phase slope as a function of the frequency at constant mass is
obtained by differentiation of Eq.~(\ref{eq:phi}):
\begin{equation} \label{eq:slopderivEMF}
\left. \frac{\partial \phi}{\partial \omega}\right|_\sigma =
-\left( \frac{D}{V_g}+\frac{L-D}{V_{g0}}\right) \left(
\frac{1+\alpha \cos (\omega \tau)}{1+2\alpha \cos(\omega
\tau)+\alpha^2} \right).
\end{equation}

We can establish a finalized equation taking into account the
electromagnetic interferences by combining
Eqs.~(\ref{eq:dphidVatom}) and (\ref{eq:slopderivEMF}) in
Eq.~(\ref{eq:phaseOmSigma2}):
\begin{equation}\label{eq:SomegaInterf}
    S_\omega = \frac{D}{V} \left( \frac{D}{V_g} +
    \frac{L-D}{V_{g0}} \right)^{-1} S_V.
\end{equation}

At the opposite of the open loop configuration, the frequency
sensitivity is not influenced by the interferences. The
perturbation caused by interferences on the aspect of the phase is
cancelled by the closed loop configuration. However, as indicated
by Eq.~(\ref{eq:SomegaInterf}), the frequency sensitivity is
strongly dependent of the structure of the sensor and the
velocities in the different parts of the sensor. A simple
expression can not be deduced easily and the link between the
frequency sensitivity and the gravimetric sensitivity is difficult
to exploit directly unless some assumptions are considered as
explained here after.

If the waveguide is not dispersive and $V=V_0$, frequency
variations obtained experimentally are directly linked to the
gravimetric sensitivity by the ratio $D/L$ as seen by replacing
the transit time obtained via Eq.~(\ref{ap:eq:F}) in
Eq.~(\ref{eq:SomegaInterf}):
\begin{equation} \label{ap:eq:domdsigma}
S_\omega = \frac{D}{L} S_V.
\end{equation}
If the waveguide is dispersive, the transit time $\tau$ contains
the combined information of the group velocities in the
transducing and sensing part and the phase velocity in the sensing
part. If the sensing part extends to the entire delay line
($D=L$), we obtain an expression corresponding to a well-known
result (for instance \cite{Jakoby97}):
\begin{equation}\label{eq:Jakoby97}
    S_\omega = \frac{V_g}{V}S_V.
\end{equation}

\section{Experimental results}
\label{experimentalresults}

For the practical consideration of the described behavior, we
investigated a Love mode sensor. It was fabricated and tested
under liquid conditions to evaluate the influence of the
electromagnetic interferences. The Love mode was obtained by
conversion of a surface skimming bulk wave (SSBW) launched in the
direction perpendicular to the crystalline X axis of a $500\
\mathrm{\mu m}$ thick ST-cut ($42.5^\circ$ Y-cut) quartz
substrate. The conversion was achieved by a $1.2\ \mathrm{\mu m}$
thick overlayer of silicon dioxide deposited on the top side of
the substrate by plasma enhanced chemical vapor deposition
(Plasmalab 100 from Oxford Plasma Technology, England). Vias were
etched in the silicon dioxide layer using a standard
$\mathrm{SF_6/O_2}$ plasma etch recipe. This process stopped
automatically on the aluminium contact pads of the transducers.

The transducers consist of split fingers electrodes etched in
$200$ nm thick sputtered aluminium. The fingers are $5\
\mathrm{\mu m}$ wide and equally spaced by $5\ \mathrm{\mu m}$.
This defines a periodicity $\lambda_0$ of $40\ \mathrm{\mu m}$.
The acoustic aperture defined by the overlap of the fingers is
equal to $80\lambda_0$ (= 3.2 mm), the total length of each IDT is
$100\lambda_0$ (= 4 mm) and the distance center to center of the
IDTs is $225 \lambda_0$ ($L$= 9 mm, $D$= 5 mm).

The sensing area was defined by covering the space left between
the edges of the IDTs by successive evaporation and lift-off of 10
nm of titanium and 50 nm of gold in a first experiment, and 200 nm
of gold in a second experiment. The fingers were protected against
liquid by patterning photosensitive epoxy SU-8 2075 (Microchem
Corp., MA) defining $200\ \mathrm{\mu m}$ thick and $80\
\mathrm{\mu m}$ wide walls around the IDTs. Quartz glasses of 5 by
5 $\mathrm{mm^2}$ were glued on top of the walls to finalize the
protection of the IDTs.

The device was mounted and wire-bonded to an epoxy printed circuit
board and its transfer function was recorded on a HP4396A Network
Analyzer. This setup corresponds to the open loop configuration.
Epoxy around the device covered and protected it and defined a
leak-free liquid cell. The sensing area was immersed in a solution
of $\mathrm{KI/I_2}$ (4 g and 1 g respectively in 160 ml of water)
that etched the gold away of the surface \cite{Vossen78}. The
transfer function of the device was recorded every 4 seconds
(limited by the GPIB transfer speed) during the etching of the
gold with a resolution of 801 points over a span of 2 MHz centered
around 123.5 MHz. The initial transfer function of the device is
presented in Figure~\ref{fig:w38d17init} with and without gold.
The transfer function during etching of the 200 nm is shown at two
moments (44 seconds and 356 seconds after etching start) in
Figure~\ref{fig:w38d17etch}. The total time for this etching was
approximately 620 seconds. The sensitivity was calculated by
etching of 50 nm of gold and assuming a density $\rho=19.3\
\mathrm{g/cm^3}$. The result is plotted versus the frequency in
Figure~\ref{fig:sens50nm}.

\section{Discussion}
\label{discussion}

Electromagnetic interferences have a clear effect on the transfer
function because of the ripples they cause. The interaction
modelled as a constant factor $\alpha$ is specific to each device
and must be identified via a careful inspection of the transfer
function. The amplitude of the transfer function peak to peak is
supposed to be the product between the transfer function of the
transducers and the interference, and therefore an evaluation of
$\alpha$ is possible if the transfer function of the transducers
only is known. However, the experiment shows that $\alpha$ is a
function of the frequency and the surface density, indicating that
finding its exact value is not straightforward. Only the phase
indicates whether $\alpha$ is higher or lower than one.

In term of sensitivity, when $\alpha \ge 1$ the phase has a
periodicity $P$ in the range $0$ to $\pi$. We suggest the
following correction to the experimental phase sensitivity:
\begin{equation}\label{eq:SphicorrectedP}
S_\phi = \frac{2\pi}{P}\frac{1}{kD}\frac{\d \phi}{\d \sigma}.
\end{equation}
This modification gives a better evaluation of the gravimetric
sensitivity by stretching the phase of the transfer function to
$2\pi$. Only the extraction of $P$ is not immediate since it
depends upon $\alpha$.

Finally, we must mention that the experimental part is not exactly
providing a differential surface density $\d \sigma$. Indeed,
etching of $50\ \mathrm{nm}$ of gold corresponds to a surface
density change of $96.5\ \mathrm{\mu g/cm^2}$. This is a
relatively large shift compared to the targeted biochemical
recognition application where protein films surface density are in
the order of $500\ \mathrm{ng/cm^2}$. The evaluation of the
sensitivity is best recorded by adding or etching thin layers of
materials and that under the operating conditions of the sensor,
especially if liquids are involved \cite{Friedt03}.

\section{Conclusion}
\label{conclusion}

We have proposed a model for surface acoustic waveguides used as
sensors. The model shows the influence of electromagnetic
interferences caused by interdigital transducers on the
gravimetric sensitivity in open and closed loop configurations. In
both cases, the dimensions of the delay line and the sensing part
influence the experimental value of phase or frequency shifts.

In an open loop configuration and with interferences, the {\em
phase} shift is disturbed and the sensitivity is over- or
under-estimated to the value of the gravimetric sensitivity. For
strong interferences, the phase has a periodicity lower than
$2\pi$ that must be considered when normalizing the phase shift to
obtain a correct figure of the sensitivity.

In a closed loop configuration and with interferences, the {\em
frequency} shift is not disturbed. The frequency shift is
proportional to the sensitivity by the ratio between the length of
the sensing area and the distance separating the transducers. In
addition, the frequency shift is influenced by the dispersive
properties of the waveguide.

The influence of the electromagnetic interferences on the transfer
function of a Love mode sensor operating in liquid conditions was
presented for a comparison. From the experiment it appears that
the interferences are function of both the frequency and the
surface density.

For future investigations, an analytical expression of the
electromagnetic-acoustic interaction and the parameters acting on
it have to be identified in order to reduce the influence or, on
the opposite, to enhance the gravimetric sensitivity of surface
acoustic waveguides.

\section{Acknowledgements}
\label{acknowledgements}

L. Francis is thankful to N. Posthuma for the support with the
PECVD tool, to C. Bartic for the SU8 walls fabrication, and to the
belgian F.R.I.A. fund for financial support.



\newpage
\begin{figure}[h!tb]
\begin{center}
\end{center}
\caption{Relative insertion loss (top) and phase (bottom) of the
transfer function for $\alpha=0$.} \label{fig:phasealpha0}
\end{figure}

\begin{figure}[h!tb]
\begin{center}
\end{center}
\caption{Relative insertion loss (top) and phase (bottom) of the
transfer function for $\alpha=1/2$.} \label{fig:phasealpha1}
\end{figure}

\begin{figure}[h!tb]
\begin{center}
\end{center}
\caption{Relative insertion loss (top) and phase (bottom) of the
transfer function for $\alpha=1$.} \label{fig:phasealpha2}
\end{figure}

\begin{figure}[h!tb]
\begin{center}
\end{center}
\caption{Relative insertion loss (top) and phase (bottom) of the
transfer function for $\alpha=2$.} \label{fig:phasealpha3}
\end{figure}

\begin{figure}[h!tb]
\begin{center}
\end{center}
\caption{Phase sensitivity at constant frequency as a function of
the relative frequency for different values of simulated
interferences obtained by Eq.~(\ref{eq:phaseSinterf}).}
\label{fig:constantfreq}
\end{figure}

\begin{figure}[h!tb]
\begin{center}
\end{center}
\caption{Initial aspect of the experimentally recorded transfer
function of the Love mode sensor with (dashed line) and without
(solid line) an overlayer of 200 nm of gold. This device presents
an initial phase $\phi_0=\pi$, leading to a vertical offset by
$\pi$ compared to the simulated phase curve represented in
Fig.~\ref{fig:phasealpha1}.} \label{fig:w38d17init}
\end{figure}

\begin{figure}[h!tb]
\begin{center}
\end{center}
\caption{Aspect of the experimentally recorded transfer function
at two different moments of the etching of 200 nm of gold (solid
line after 44 seconds and dashed line after 356 seconds). The
solid line shows a value of $\alpha$ close to 1 around $123.5\
\mathrm{MHz}$.} \label{fig:w38d17etch}
\end{figure}

\begin{figure}[h!tb]
\begin{center}
\end{center}
\caption{Phase sensitivity computed with help of the experimental
data obtained from etching of 50 nm of gold as a function of the
frequency. Oscillations are attributed to electromagnetic
interferences.} \label{fig:sens50nm}
\end{figure}

\vfill

\clearpage
\includegraphics[width=12cm]{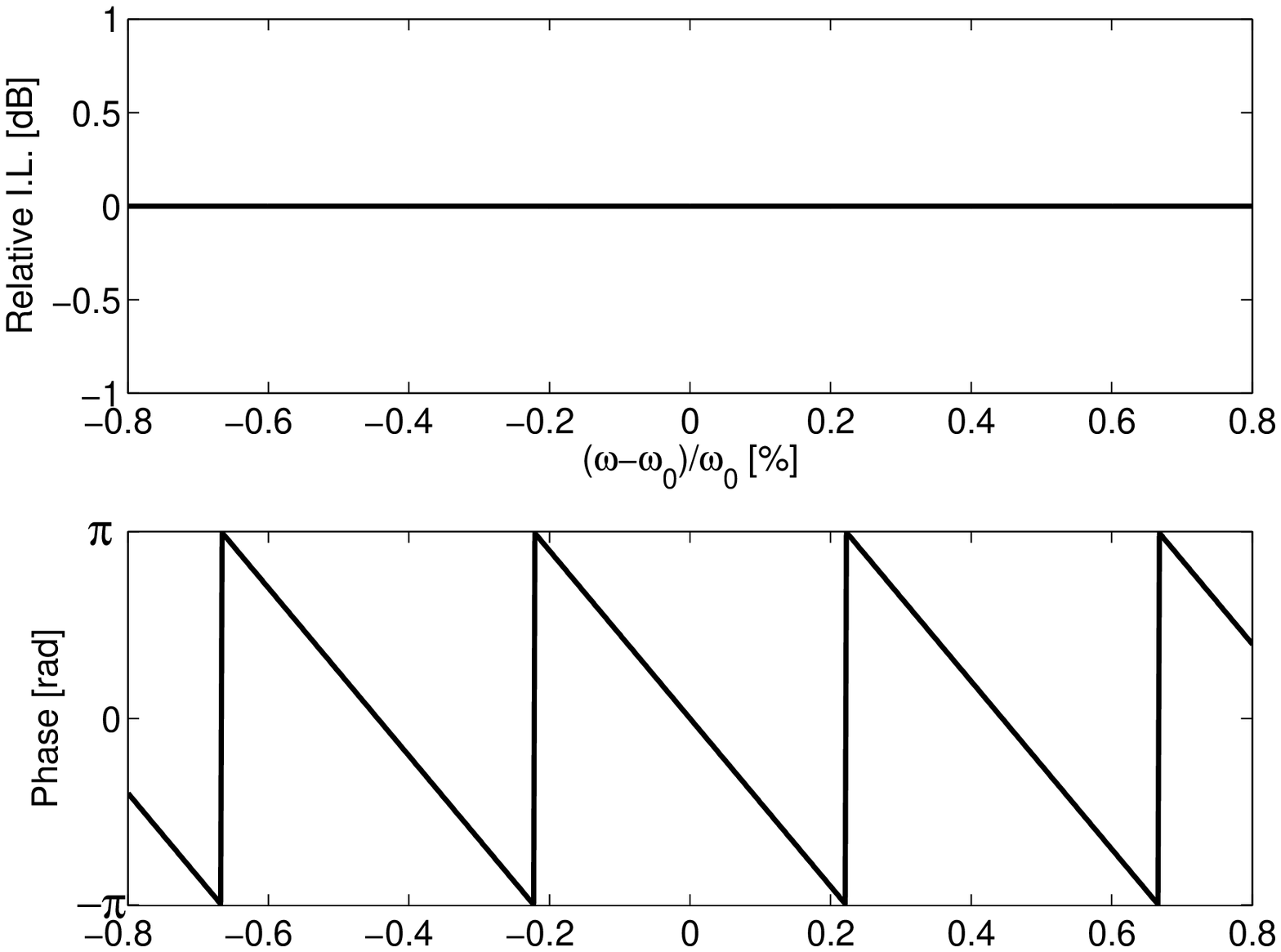}

\vfill Figure 1

\newpage
\includegraphics[width=12cm]{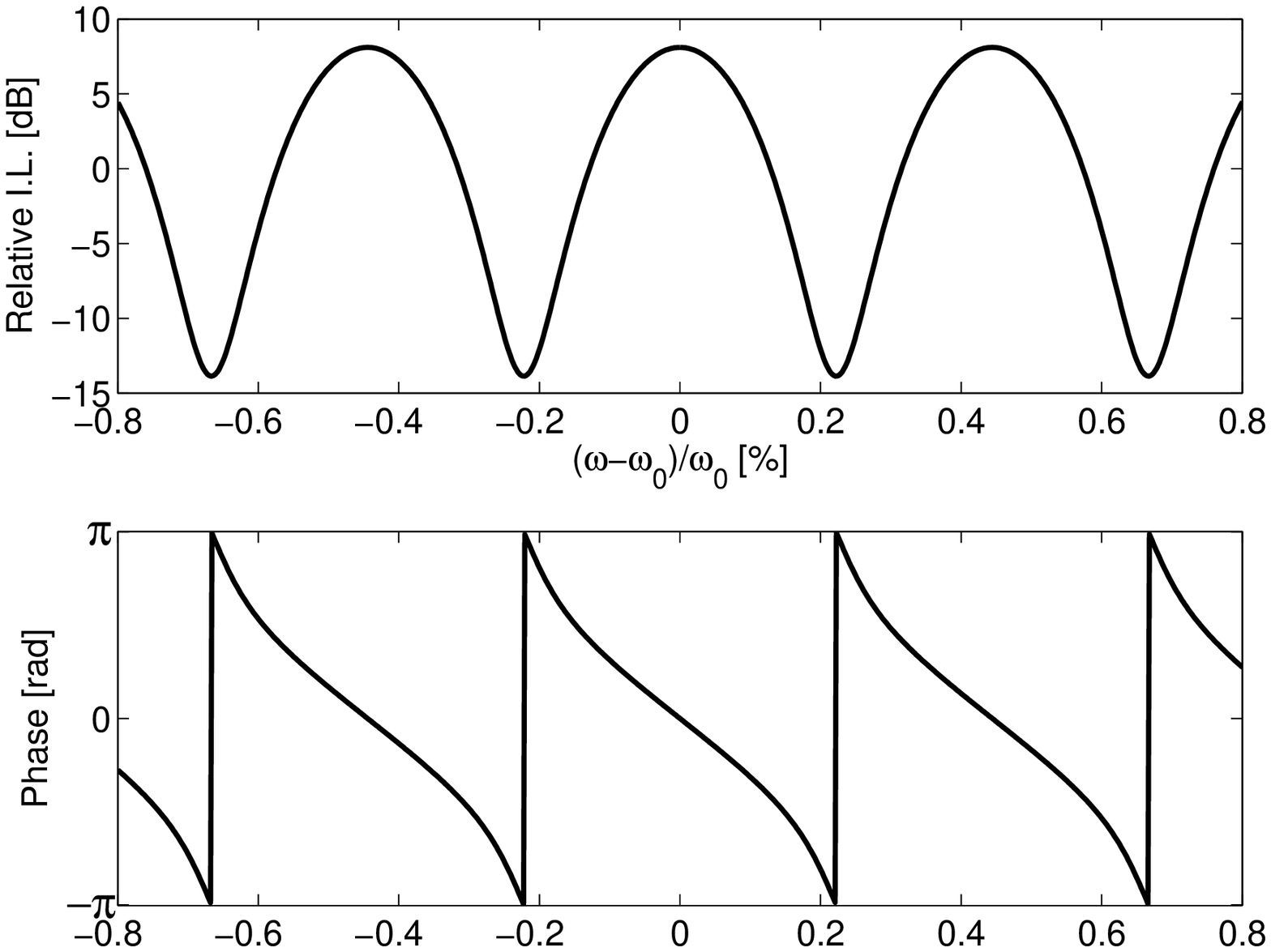}

\vfill Figure 2

\newpage
\includegraphics[width=12cm]{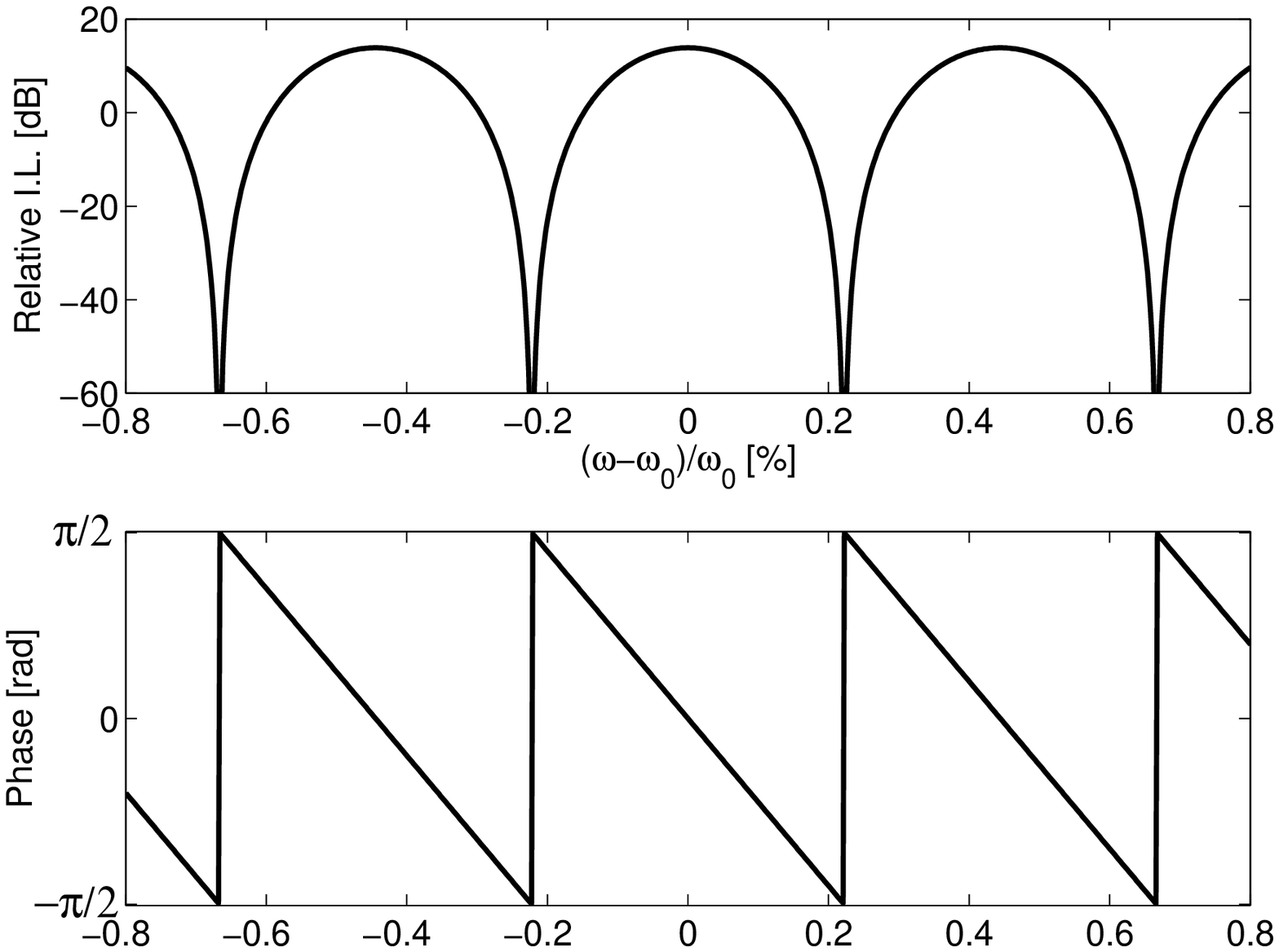}

\vfill Figure 3

\newpage
\includegraphics[width=12cm]{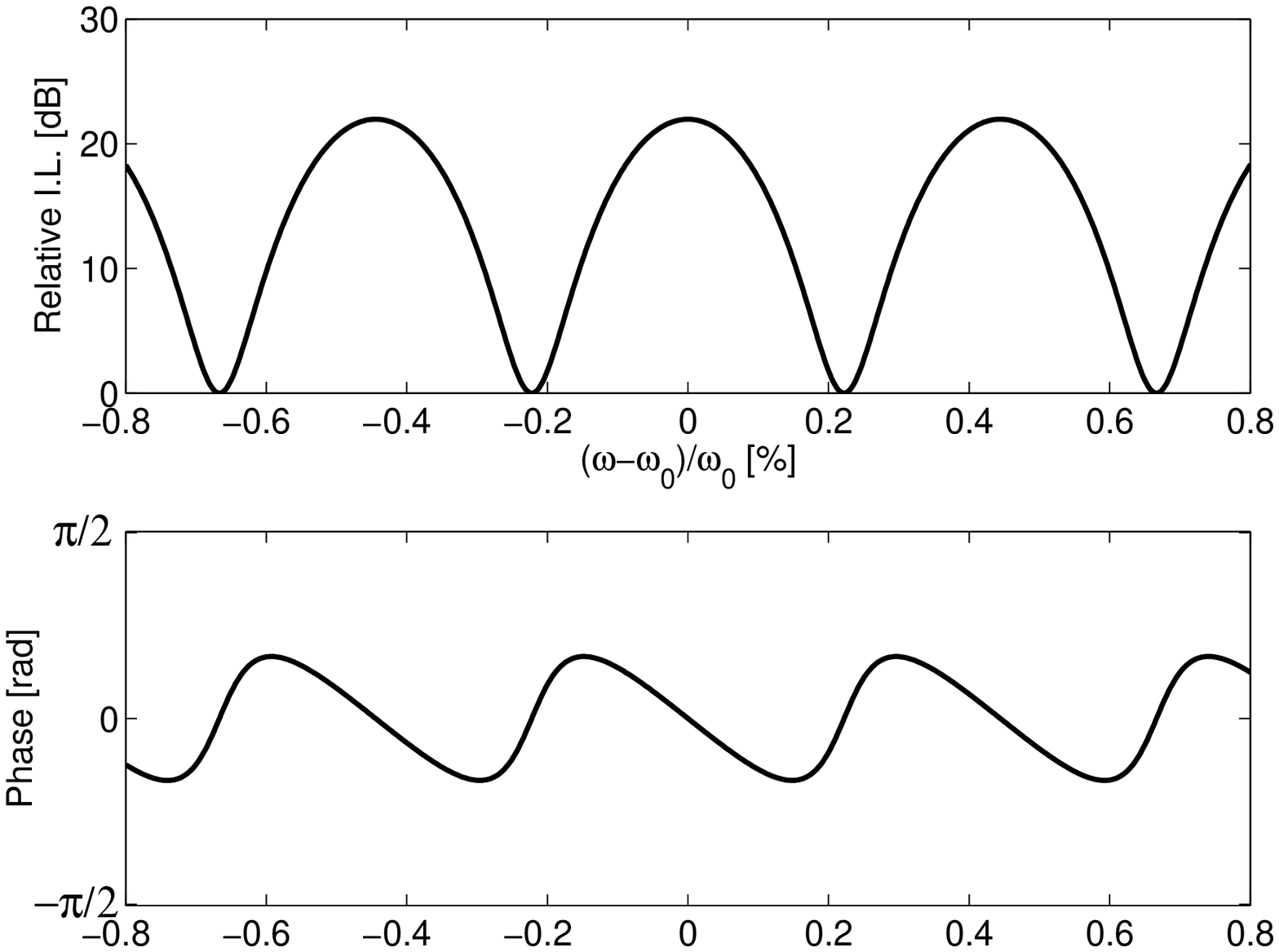}

\vfill Figure 4

\newpage
\includegraphics[width=12cm]{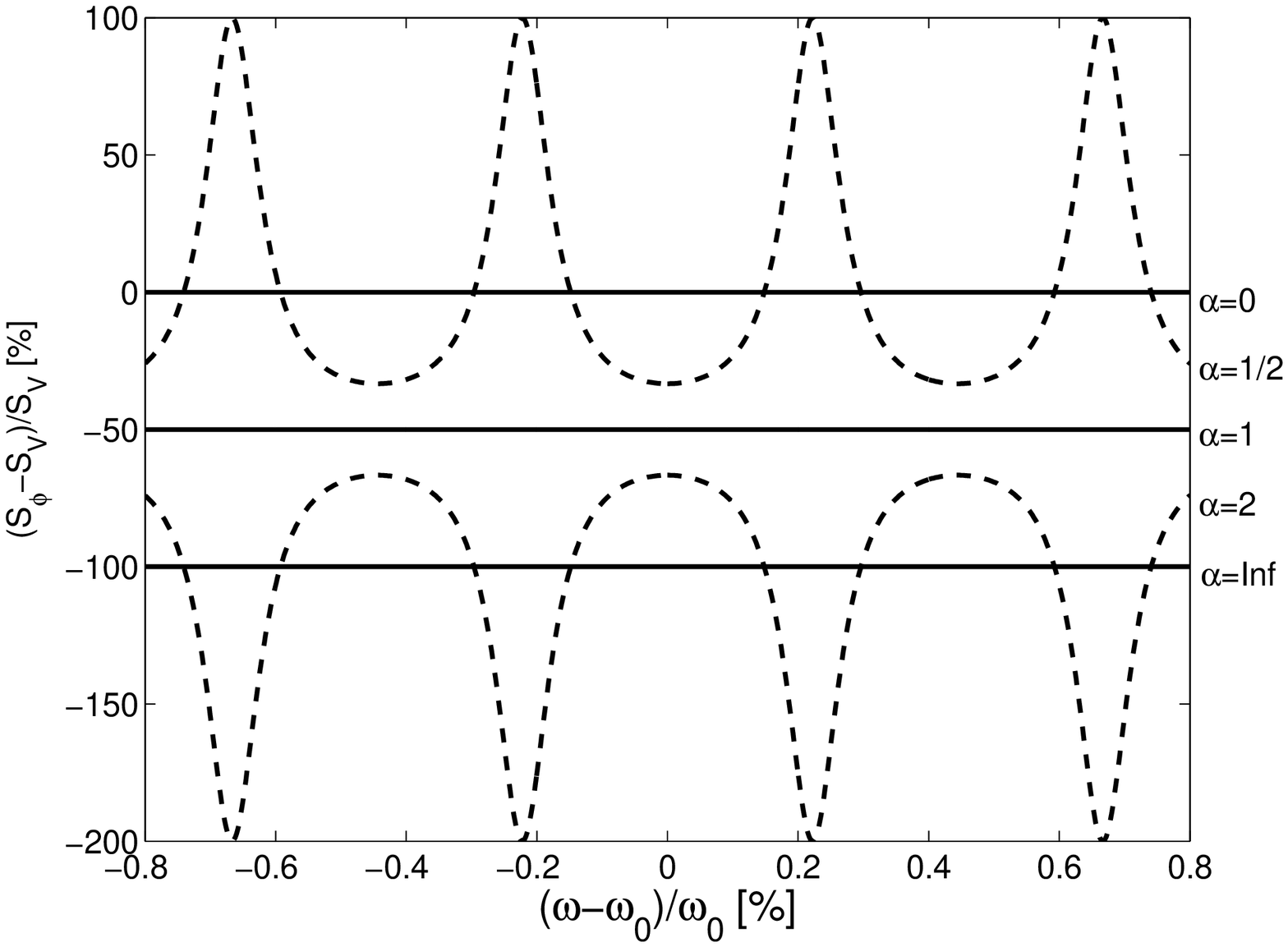}

\vfill Figure 5

\newpage
\includegraphics[width=12cm]{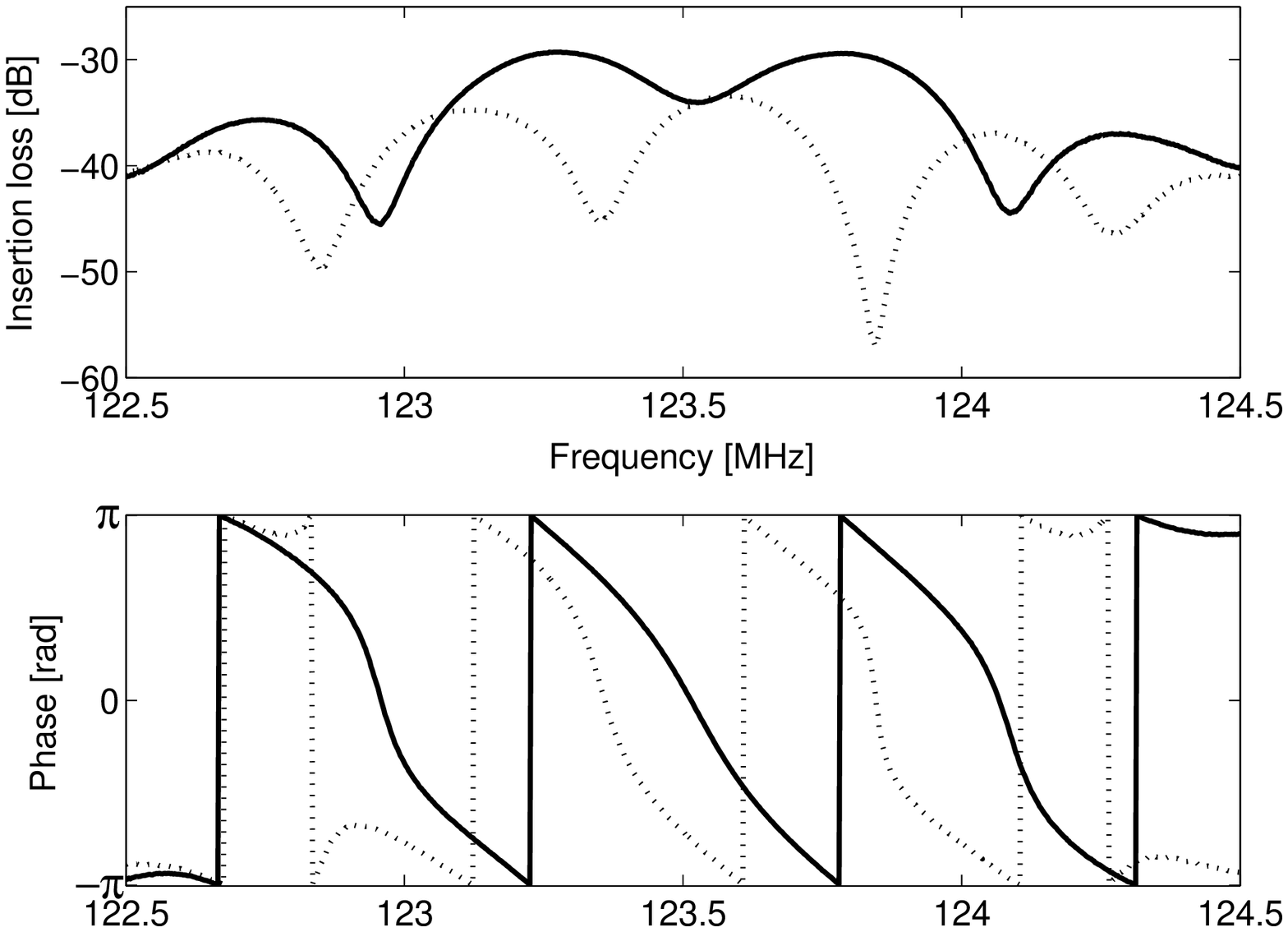}

\vfill Figure 6

\newpage
\includegraphics[width=12cm]{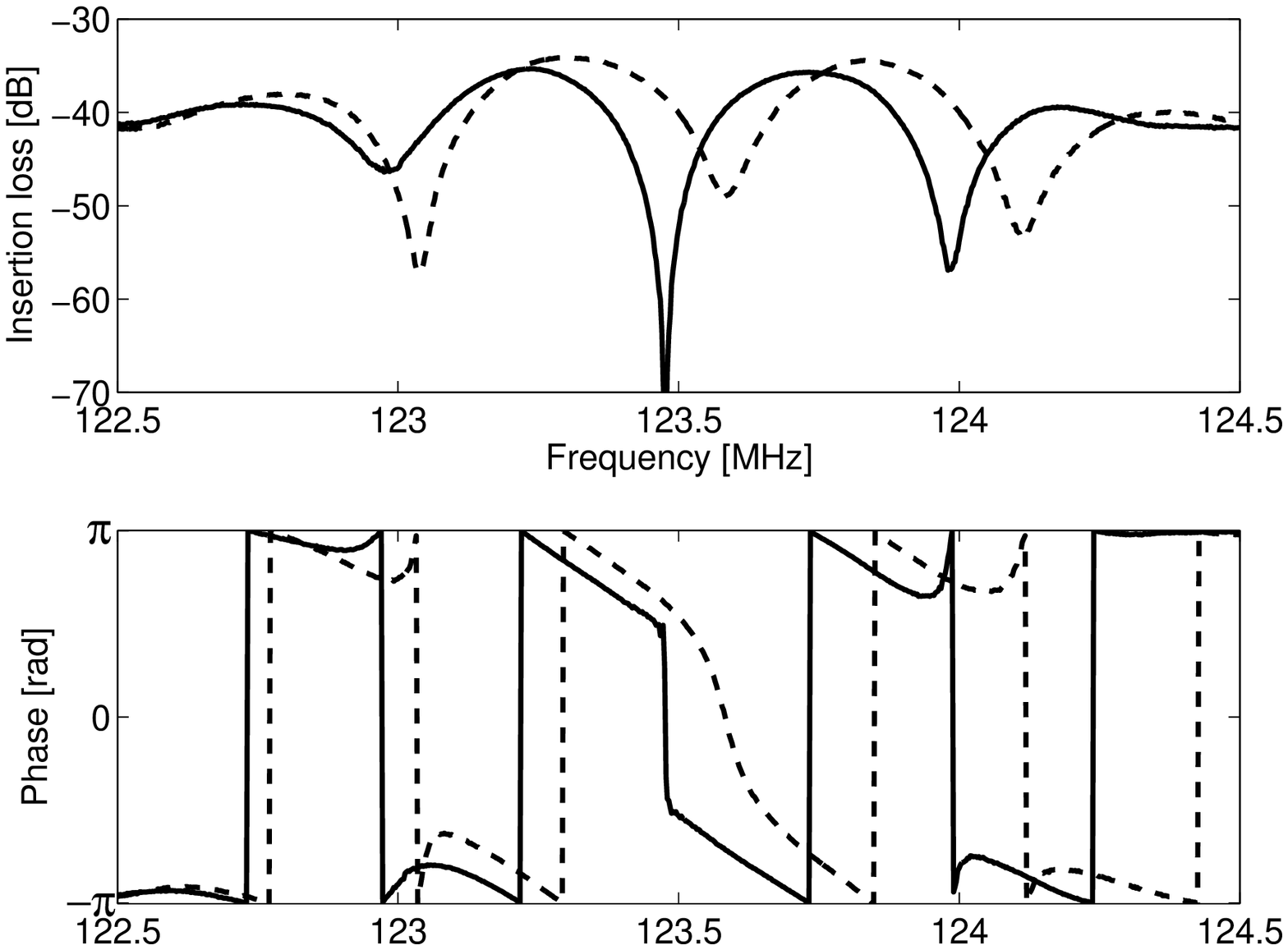}

\vfill Figure 7

\newpage
\includegraphics[width=12cm]{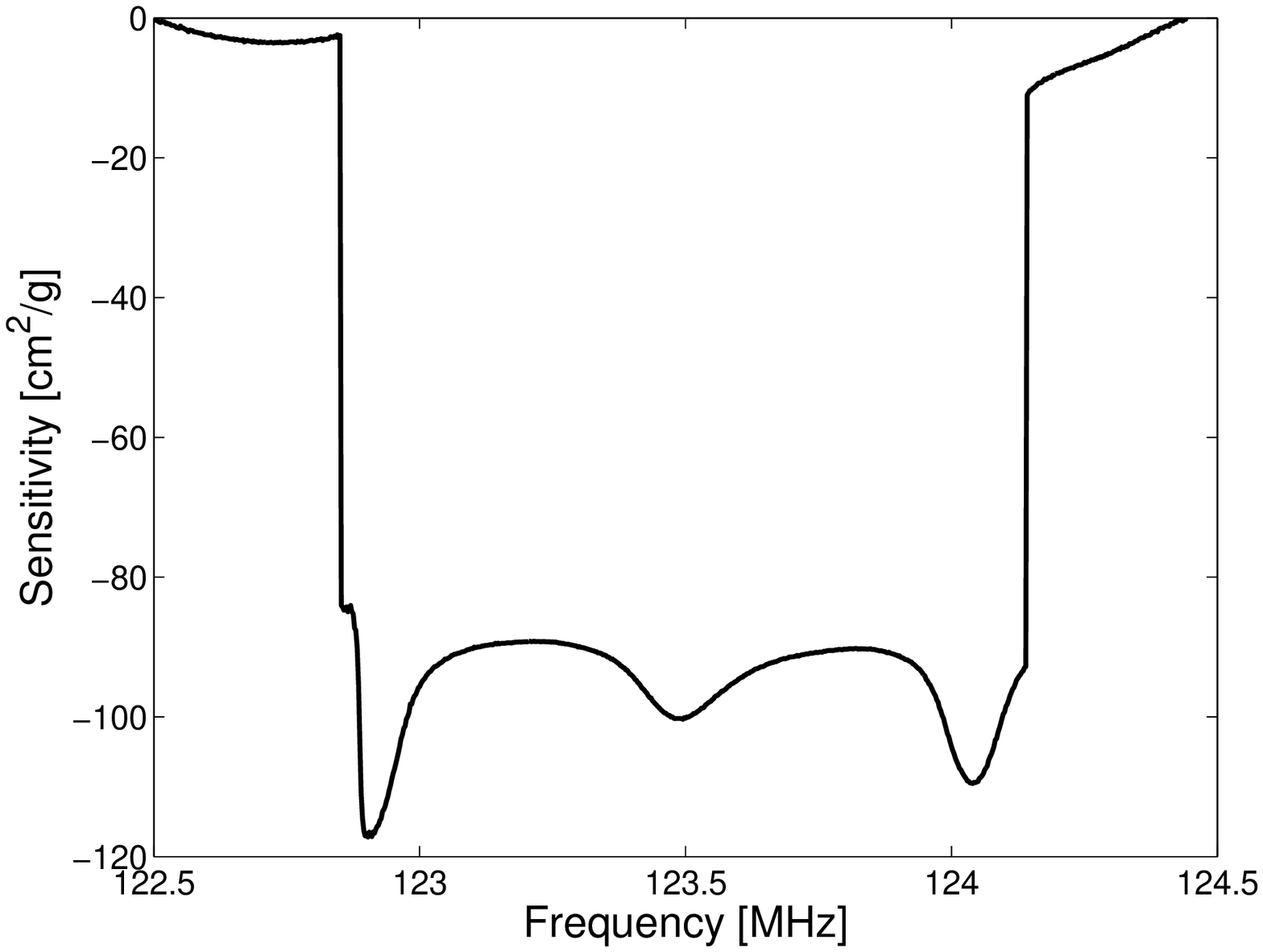}

\vfill Figure 8

\end{document}